\documentclass[12pt]{article}
\usepackage{epsfig,ldd_art,fullpage}
\newcommand{\lapprox}{\raisebox{-0.5ex}{$\
\stackrel{\textstyle<}{\textstyle\sim}\ $}}

\usepackage{graphicx}
\newcommand{\be}{\begin{equation}}
\newcommand{\ee}{\end{equation}}
\newcommand{\bea}{\begin{eqnarray}}
\newcommand{\eea}{\end{eqnarray}}

\newcommand{\One}{1\kern-4.5pt1}
\begin{document}
\begin{center}

\begin{flushright}
     SWAT/02/341 \\
August 2002
\end{flushright}
\par \vskip 10mm

\vskip 1.2in

\begin{center}
{\LARGE\bf
Application of the \\Maximum Entropy Method 
\vskip 0.08in
to the $(2+1)d$ Four-Fermion Model }

\vskip 0.7in
C.R. Allton $\!^{a,b}$, J.E. Clowser $\!^a$, S.J. Hands $\!^a$, 
J.B. Kogut $\! ^c$ and C.G. Strouthos $\!^{a}$ \\
\vskip 0.2in

$^a\,${\it Department of Physics, University of Wales Swansea,\\
Singleton Park, Swansea, SA2 8PP, U.K.} \\
$^b\,${\it Department of Mathematics, University of Queensland,\\
Brisbane 4072, Australia.} \\
$^c\,${\it Department of Physics, University of Illinois at Urbana-Champaign,\\
Urbana, Illinois 61801-3080, U.S.A.}\\
\end{center}

\vskip 0.5in 
{\large\bf Abstract}
\end{center}
\noindent
We investigate spectral functions 
extracted using the Maximum Entropy Method   
from correlators measured in lattice simulations of the (2+1)-dimensional 
four-fermion model.
This model is particularly interesting because it has both
a chirally 
broken phase with a rich spectrum of mesonic bound states and a symmetric 
phase where there are only resonances. 
In the broken phase we study the 
elementary fermion, pion, sigma, and massive pseudoscalar meson; our results
confirm the Goldstone nature of the $\pi$ and permit an estimate of the meson
binding energy. We have, however, seen no signal of $\sigma\to\pi\pi$ decay as
the chiral limit is approached.
In the
symmetric phase we observe a resonance of non-zero width in qualitative
agreement with analytic expectations; in addition the ultra-violet behaviour of 
the spectral functions is consistent with the large non-perturbative
anomalous dimension for fermion composite operators expected in this model.

\newpage

\section{Introduction}

The Gross-Neveu model in $d=3$ spacetime dimensions (GNM$_3$) has been the
object of much analytic and numerical study in recent years.
Its Lagrangian density is 
\begin{eqnarray}
{\cal L}&=& \bar{\psi}_i(\partial\hskip -.5em / + m) \psi_i
- \frac{g^2}{2 N_f}[(\bar{\psi}_i \psi_i)^2-(\bar{\psi}_i \gamma_5 \psi_i)^2]\cr
&\sim&
\bar{\psi}_i\biggl(\partial\hskip -.5em / + m + {g\over\surd N_f}
(\sigma + i \gamma_5 \pi)\biggr)
\psi_i
+ \frac{1}{2} (\sigma^{2}+ \pi^2),
\label{eq:L}
\end{eqnarray}
where the index $i$ runs over $N_f$ fermion flavors and in the second line
we have introduced scalar $\sigma$ and pseudoscalar $\pi$ auxiliary boson
fields.
Apart from the obvious numerical advantages of working with a 
relatively simple theory in a reduced
dimensionality
there are several features which make GNM$_3$ interesting for the modelling of 
strong interactions \cite{rosen91}. 

\begin{itemize}

\item
For sufficiently 
strong coupling $g^2>g_c^2$ it exhibits spontaneous chiral symmetry breaking 
implying dynamical generation of a fermion mass $M_f$, 
the pion field $\pi$ 
being the associated Goldstone boson.  A separation of scales $m_\pi\ll M_f$
is possible.

\item 
The spectrum 
of excitations contains both ``baryons'' and ``mesons'', 
namely the elementary fermions $f$ and
the composite $f\bar f$ states; 

\item
For $2<d<4$ there is
an interacting continuum limit
at a critical value of the coupling, which for $d=3$ has a numerical value 
$g_c^2/a\approx1.0$ in the large-$N_f$ limit if a lattice
regularisation is employed \cite{kogut93}. There is a renormalisation group UV
fixed point at $g^2=g_c^2$, signalled by the
renormalisability of the $1/N_f$ expansion \cite{rosen91}, 
entirely analagous to the Wilson-Fisher fixed point 
in scalar field theory.

\item
Numerical simulations 
with baryon chemical potential $\mu\not=0$ 
show qualitatively correct behaviour, in that the onset of matter occurs
for $\mu$ of the same order as
the constituent quark scale $M_f$ \cite{hands95}, 
rather than for $\mu\approx m_\pi/2$,
which happens in gauge theory simulations with a real measure
$\mbox{det}(M^\dagger M)$ because of the presence of 
a baryonic pion in the spectrum.
This makes GNM$_3$ an ideal arena in which to test 
strongly interacting thermodynamics \cite{general}.

\end{itemize}

Let us briefly review the physical content of the model as predicted by the
large-$N_f$ approach \cite{rosen91,kogut93}. For $g^2>g_c^2$ the fermion 
has a dynamically generated mass $M_f$ given, up to corrections of 
order $1/N_f$, by
\begin{equation}
M_f={g\over\surd N_f}\langle\sigma\rangle
={g^2\over N_f}\langle\bar\psi\psi\rangle. 
\end{equation}
Its inverse 
defines a correlation length which diverges as $(g^2-g_c^2)^{-\nu}$
with critical index $\nu=1+O(1/N_f)$. 
In addition as a result
of $f\bar f$ loop corrections
the $\sigma$ and $\pi$ fields acquire non-trivial dynamics, the inverse
$\sigma$ propagator being given as a function of $d$ 
to leading order in $1/N_f$ by
\begin{equation}
D_\sigma(k^2)={1\over
g^2}{{(4\pi)^{d\over2}}\over{2\Gamma(2-{\textstyle{d\over2}})}}
{M_f^{4-d}\over{(k^2+4M_f^2)F(1,2-{\textstyle{d\over2}};{3\over2};
-{k^2\over{4M_f^2}})}}.
\label{eq:Dsigma}
\end{equation} 
Immediately we see the difference between this model 
and QCD.
For $k^2\ll M_f^2$ $F\approx1$, implying that
to this order the $\sigma$ resembles a weakly-bound meson of mass 
$M_\sigma=2M_f$; however,
the hypergeometric function 
in the denominator indicates a strongly interacting $f\bar f$
continuum immediately above the threshold $2M_f$.
This implies that if truly bound, its binding energy is
$O(1/N_f)$ at best 
(to our knowledge there have so far been no analytic calculations), implying
little if any separation between pole and threshold.
Since all residual interactions 
are subleading in $1/N_f$, we surmise that all other mesons 
are similarly weakly bound states of massive fermions, and hence effectively
described by a two-dimensional ``non-relativistic quark model''.
A recent study of mesonic wavefunctions in GNM$_3$ provides 
evidence for this picture \cite{HKS}.
In an asymptotically-free but confining theory like QCD, by contrast, 
one expects isolated poles and/or resonances,
corresponding to relativistic bound states in the channel in question, 
which 
are well separated from a threshold to nearly-free quark behaviour
which sets in 
at typically 1.3 - 1.5 GeV \cite{shuryak}.

The exception to this rule is the pion.
The Lagrangian (\ref{eq:L}) can be defined with either a continuous U(1)
or discrete Z$_2$ chiral symmetry, 
the latter case being realised by setting the
$\pi$ field to zero. In the case of U(1) chiral symmetry, 
for $m=0$ and $g^2>g_c^2$ 
the pion propagator $D_\pi$ is given by a similar expression 
to (\ref{eq:Dsigma})
with the factor $(k^2+4M_f^2)$ replaced by $k^2$; the massless
pole demonstrates that $\pi$ couples to a Goldstone mode. 
For $m>0$, we expect by the usual PCAC arguments that the $\pi$ acquires
a mass $m_\pi\propto\surd m$, and that the ratio $m_\pi/M_f$ can be tuned
to be arbitrarily small. In particular, once it is less than unity the $\sigma$
becomes unstable with respect to decay into $2\pi$.
Note, however, that
the Goldstone mechanism in GNM$_3$ is fundamentally different
from that in QCD. 
In GNM$_3$ the diagrams responsible for making the 
pion light are flavor-singlet
chains of disconnected $f\bar f$ bubbles \cite{hands95}. 
The non-singlet connected $f\bar f$ diagram which interpolates the pion in QCD 
corresponds in GNM$_3$ to a pseudoscalar state
with mass $O(2M_f)$.

For $g^2<g_c^2$ the model 
is chirally 
symmetric, and hence all states are massless, as $m\to0$. In this limit 
$D_\sigma$ and $D_\pi$ coincide, and 
in the large-$N_f$ limit 
neither has a pole on the physical sheet \cite{rosen91}.
The auxiliary fields in this case do not interpolate to a stable particle. 
A dimensionful scale is still defined, however, 
by the width $\mu$ of a resonance in
$f\bar f$ scattering in these channels; this diverges as $(g^2_c-g^2)^{-\nu}$
with the same exponent $\nu$ \cite{kogut93}. 

It is clear that despite its simplicity GNM$_3$ exhibits 
phenomena such as resonances, decays and multi-particle continua which 
are not easily analysed using the traditional 
techniques of single- and multi-exponential
fitting to Euclidean correlators developed for quenched QCD.
This was recognised in early studies, which attempted fits inspired by the
large-$N_f$ forms of $D_{\sigma}$ in both chirally broken and symmetric
phases, with ambiguous results \cite{kogut93}.
A more systematic approach, however,  is to focus on the {\em spectral density
function\/} $\rho(\omega)$, defined implicitly via the Euclidean timeslice
meson correlator $C(t)$ by 
\begin{equation}
C(t)=\sum_{\vec x}\langle J(\vec 0,0)J^\dagger(\vec x,t)
\rangle=\int_0^\infty d\omega
\rho(\omega)e^{-\omega t}.
\label{eq:defrho}
\end{equation}
Here $J$ is a local fermion bilinear $\bar\psi\Gamma\psi$ which in principle
projects onto all physical states consistent with 
a given set of quantum numbers. All information about bound states, resonances
and particle production thresholds as a function of energy $\omega$
is contained in $\rho$. The procedures for fitting lattice-generated
data to date have assumed {\em Ans\"atze} for $\rho$ such as one or more 
bound state poles of the form $\delta(\omega-M)$, or perhaps a free 
particle continuum above some threshold \cite{QCDSR}. However, more recent
works have attempted {\em ab initio} calculations of $\rho(\omega)$
\cite{hatsuda,MEMlat,FLPSW}. 
This is a difficult problem: the inversion of (\ref{eq:defrho})
is ill-posed since $\rho(\omega)$ is a continuous function whereas 
lattice simulations only yield $C(t)$ for a discrete, finite set of points, and
moreover with some statistical uncertainty. The approach adopted in
Refs.~\cite{MEMlat} is to apply the {\em Maximum Entropy Method} (MEM) which 
attempts to fit $\rho(\omega)$ subject to reasonable assumptions of smoothness
and stability with respect to small variations in the input data.

In this paper we present results from a study of spectral functions 
extracted from numerical simulations of GNM$_3$ using MEM techniques.
To our knowledge this is the first such study beyond the quenched approximation.
Our goal is to explore some of the features described above which distinguish
GNM$_3$ from quenched QCD. In this regard it is worth noting that because the 
two most important mesonic channels, $\sigma$ and $\pi$, 
are represented by bosonic auxiliary fields,
the correlation functions in these channels automatically
include the disconnected
diagrams which are so expensive to calculate in QCD; 
in GNM$_3$, by contrast, these can be measured with high statistics
relatively cheaply. 
We will also examine the fermion and non-singlet pseudoscalar (PS) channels.
As surveyed above, simulations of GNM$_3$ offer the freedom to vary the 
phase of the theory (by varying $\mbox{sgn}(g^2-g_c^2)$), the 
correlation length (by varying $\vert g^2-g_c^2\vert$), the symmetry group
(by including or omitting $\pi$), the ratio $m_\pi/M_f$ (by varying
$m$), and the interaction strength (by varying $N_f$) -- 
in the current study we will exploit most of these opportunities. In future 
work we plan also to study the model with both non-zero 
temperature $T$ and baryon
chemical potential $\mu$.

In Sec.~\ref{sec:mem} we survey MEM and explain our
implementation of it. Sec.~\ref{sec:theory} outlines some theoretical 
expectations related to $\rho(\omega)$ in GNM$_3$ based on the large-$N_f$
approach, and Sec.~\ref{sec:sim} details the lattice formulation 
and numerical simulations. Our results are presented in Sec.~\ref{sec:results},
and conclusions in Sec.~\ref{sec:summary}.

\section{The Maximum Entropy Method}
\label{sec:mem}

The theoretical basis for MEM is
Bayes' theorem in probability theory \cite{PROB.98.HJ}:
\be
  \label{eq:bayes}
  P[X|Y]=\frac{P[Y|X]P[X]}{P[Y]},
\ee
where $P[X|Y]$ denotes the conditional probability of $X$
given $Y$. In terms of the lattice data $D$,
spectral function $\rho$ and all {\em a priori} knowledge $H$,
Bayes' theorem reads:
\be
  \label{eq:lat_bayes}
  P[\rho|DH] = \frac{P[D|\rho H]P[\rho|H]}{P[D|H]}.
\ee
$P[D|\rho H]$ is known as the {\em likelihood function} and is
the equivalent of the familiar $\chi^2$ in the least squares method
\cite{PROB.83.SB}. For a large number of Monte Carlo measurements
of a correlation function, the data $D$ are expected to obey a gaussian
distribution according to the central limit theorem,
\bea
  \label{eq:def_likelihood}
  P[D|\rho H] & = & \frac{1}{Z_L} e^{-L[\rho]}, \\
  \label{eq:def_L}
  L[\rho] & = & \frac{1}{2} \sum_{i,j=1}^{N_t}(D(t_i) - D_\rho(t_i))
          C_{ij}^{-1} (D(t_j) - D_\rho(t_j)),
\eea
where the normalisation factor $Z_L =
(2\pi)^\frac{N_t}{2}\sqrt{\det C}$ and $N_{t}$ is the number
of temporal points. Lattice data averaged over $N_{\rm cfg}$ gauge
configurations $D(t)$, the covariance matrix $C_{ij}$, and the
propagator constructed from the spectral function $\rho$ using the 
lattice kernel $K(\omega,t)$ are defined by
\bea
  \label{eq:def_central_value}
  D(t_i) & = & \frac{1}{N_{\rm cfg}}\sum_{m=1}^{N_{\rm cfg}}
  D^m(t_i), \\
  \label{eq:def_covariance}
  C_{ij} & = & \frac{1}{N_{\rm cfg}(N_{\rm cfg}-1)}\sum_{m=1}^{N_{\rm cfg}}
           (D^m(t_i) - D(t_i))(D^m(t_j) - D(t_j)), \\
  \label{eq:spectral2corr}
  D_\rho(t) & = & \int_0^\infty K(t,\omega)\rho(\omega)d\omega.
\eea
In all our work we use a lattice kernel defined as $\exp(-\omega t)$.

The factor
$P[\rho|H]$ appearing in the numerator of (\ref{eq:lat_bayes})
is the {\em prior probability}, which is written in terms
of the Shannon-Jaynes entropy $S[\rho]$
\cite{MEM.89.JS} 
for a given {\em
default model} $\rho_0(\omega)$. The default model is usually chosen to be the
spectral function for a non-interacting two-particle continuum; for meson
states we have $\rho_0(\omega)\propto\omega^{d-2}$ (see Sec.~\ref{sec:theory}).
The final result, however, should be insensitive to the choice
of $\rho_0$.
The entropy $S[\rho]\leq0$ and becomes zero only 
when $\rho(\omega)=\rho_0(\omega)$:
\bea
  \label{eq:def_prior}
  P[\rho|H\alpha\rho_0] & = & \frac{1}{Z_S(\alpha)} e^{\alpha S[\rho]}, \\
  \label{eq:def_shannon_jaynes}
  S[\rho] & = & \int_0^\infty \left [\rho(\omega) - \rho_0(\omega)
      - \rho(\omega)\ln \left(\frac{\rho(\omega)}{\rho_0(\omega)}\right)
      \right ] d\omega \\
  \label{eq:def_discrete_shannon_jaynes}
  & \rightarrow & \sum_{\ell=1}^{N_\omega} \left [ \rho_\ell - \rho_{0\ell}
 - \rho_\ell
      \ln \left(\frac{\rho_\ell}{\rho_{0\ell}}\right) \right ]\Delta\omega,
\eea
where Eq.~(\ref{eq:def_discrete_shannon_jaynes}) results from discretising
the $\omega$-axis into $N_\omega$ bins of width $\Delta\omega$, and 
the normalisation factor $  Z_S \cong
(\frac{2\pi}{\alpha})^{\frac{N_\omega}{2}}$.
Note that two extra
parameters previously implicit in $H$ have been written in
explicitly; $\alpha$ is a real positive parameter and $\rho_0(\omega)$
a real positive function. This form of entropy
leads to a positive semi-definite spectral function in MEM.
In our work we use $N_\omega=600$ and $\Delta\omega a=0.01$.

Combining Eqs.~(\ref{eq:def_likelihood}) and (\ref{eq:def_prior})
gives
\bea
  \label{eq:step1}
  P[\rho|DH\alpha\rho_0] & \propto & \frac{1}{Z_S Z_L} e^{Q[\rho]}, \\
  \label{eq:def_Q}
  Q & \equiv & \alpha S - L.
\eea
and the condition satisfied by the most probable
spectral function $\rho_\alpha(\omega)$ is
\be
  \label{eq:max_cond}
  \left .\frac{\delta Q}{\delta \rho(\omega)} \right |_{\rho=\rho_\alpha} = 0
\ee
The parameter $\alpha$ is in effect a relative weighting
between the entropy $S$ and the likelihood $L$, and there are three
different ways
to deal with it. The value $\alpha=\hat\alpha$ can be chosen
which either gives $\chi^2 = N_{t}$ or maximises
$P[\alpha|DH\rho_0]$; these methods are known as {\em classic} and {\em
historic} \cite{MEM.89.JS} respectively. Alternatively,
a weighted
average over $P[\alpha|DH\rho_0]$ can be performed;
this is known as {\em Bryan's method} 
\cite{MEM.90.RKB} and is the one we adopt:
\be
  \label{eq:fout}
  \rho_{out}(\omega) = \int_{\alpha_{min}}^{\alpha_{max}} 
d\alpha \rho_\alpha(\omega)P[\alpha|DH\rho_0],
\ee
where $\alpha_{min}$ and $\alpha_{max}$ are chosen to 
satisfy 
\be
  \label{eq:average_criterion}
  P[\alpha_{min,max}|DH\rho_0] =0.01P[\hat{\alpha}|DH\rho_0].
\ee

\subsection{Testing MEM}
\label{subsec:testing}

To test our implementation of MEM, 
we studied an idealised QCD spectral function in the charged
$\rho$-meson channel \cite{shuryak,hatsuda}:
\be
  \label{eq:realistic_rho}
  {\rho_{in}(\omega)\over\omega^2} = \frac{2}{\pi} \left[ F_\rho^2
    \frac{\Gamma_\rho m_\rho}{(\omega^2-m_\rho^2)^2 +
    \Gamma_\rho^2 m_\rho^2} + \frac{1}{8\pi}
    (1+\frac{\alpha_s}{\pi})
    \frac{1}{1+e^{(\omega_0-\omega)/\delta}}\right],
\ee
where the pole residue $F_\rho=f_\rho m_\rho$ is defined by
\be
  \label{eq:pole_residue}
  \langle 0|\bar{d}\gamma_\mu u|\rho\rangle 
  \epsilon_\mu = \sqrt{2}f_\rho m_\rho^2
  \epsilon_\mu,
\ee
$\epsilon_\mu$ being the polarisation vector.
The following energy-dependent width is chosen with a
$\theta$-function included to give the correct threshold behaviour
of the $\rho\rightarrow\pi\pi$ decay
\be
  \label{eq:width}
  \Gamma_\rho(\omega) = \frac{g_{\rho\pi\pi}^2}{48\pi} m_\rho
  \left(1-\frac{4m_\pi^2}{\omega^2}\right)^{3/2} \theta(\omega-2m_\pi).
\ee
The values of the parameters input into (\ref{eq:realistic_rho})
are taken to be:
\bea
  \label{eq:parameter_values}
  m_\rho = 0.77,\;\; m_\pi& =& 0.14,\;\; \omega_0 = 1.3,\cr
 g_{\rho\pi\pi}=5.45,\;\; f_\rho &=& 0.184,\;\;\delta = 0.2,\;\; \alpha_s = 0.3.
\eea
where the numerical values of the first three parameters are in GeV.

\begin{figure}[htb]
\begin{center}
\epsfig{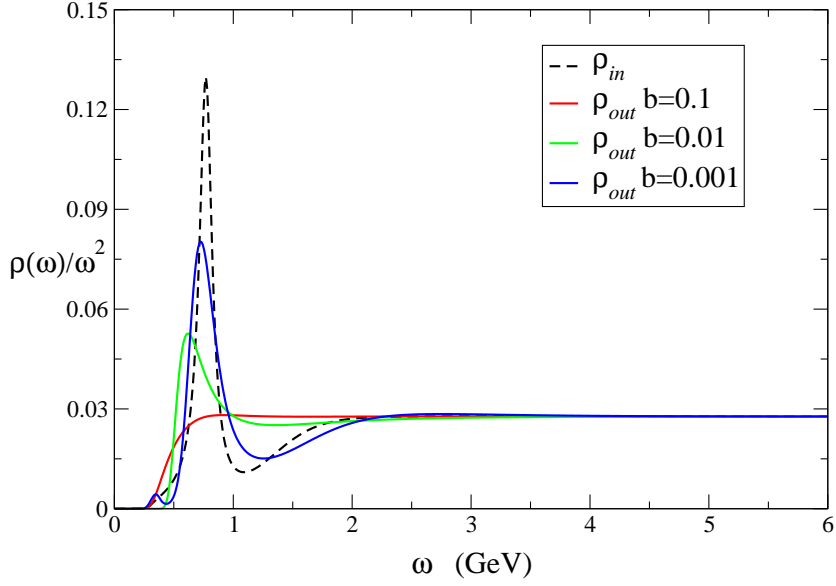}
\end{center}
\caption{Comparison of $\rho_{in}$ 
and $\rho_{out}$ for idealised data in the $\rho$-channel with $N_t=32$.}
\label{fig:testing_mem}
\end{figure}
Test lattice correlator data were constructed from the spectral function using
(\ref{eq:spectral2corr}). Gaussian noise with variance
$\sigma(t_i) = b D_{in}(t_i)t_i$ was added to this data to simulate the 
effect of decreasing signal-to-noise ratio with temporal separation
\cite{hatsuda}.
For simplicity we use a diagonal covariance matrix,
which thus neglects correlations between different $t$. The
default model used is $\rho_0(\omega) = m_0\omega^2$, 
motivated by the asymptotic behaviour of $\rho_{in}$. The parameter
$m_0$ is chosen
to be $\lim_{\omega\to\infty}\rho_{in}(\omega)=0.0277$.
We set $\omega_{max} = 6$GeV, $\Delta\omega = 10$MeV
and $N_\omega = 600$, and vary the noise parameter $b$ from 0.1
to 0.001.
Fig.~\ref{fig:testing_mem} shows a comparison between $\rho_{in}$ and
$\rho_{out}$ for various $b$.
As expected, decreasing $b$ leads
to a better agreement between input and output
spectral functions.

\section{Theoretical Preliminaries}
\label{sec:theory}

Our main focus will be 
the mesonic Euclidean timeslice correlation functions 
defined in Eqn.~(\ref{eq:defrho}).
With this definition,
if $J$ couples to a stable (ie. zero width) bound
state of mass $M$ with strength $A$ (ie.
$\langle 0\vert J\vert \vec k,M\rangle=A$), 
then $\rho(\omega)=(\vert A\vert^2/2M)\delta(\omega-M)$.
Since in $d$ spacetime
dimensions the engineering dimension $[J]=d-1$ 
and $[\vert\vec k,M\rangle]=1-{d\over2}$, 
it is readily checked that the combination
$\rho(\omega)/\omega^{d-2}$ is dimensionless. This also motivates the use of the
default model $\rho_0(\omega)\propto\omega^{d-2}$, which corresponds in
configuration space to the propagation of free massless fermions; ie. 
$C(t)\propto t^{-(d-1)}$. For an asymptotically-free theory such as QCD
we expect $\lim_{\omega\to\infty}\rho(\omega)=\rho_0(\omega)$, as
illustrated in Fig.~\ref{fig:testing_mem}: however since
GNM$_3$'s UV behaviour is described by a renormalisation group fixed point
with non-vanishing interaction strength \cite{rosen91,kogut93} this is not a
constraint in the current study.

The asymptotic form of $\rho(\omega)$ is easiest to analyse in the symmetric 
phase $g^2<g_c^2$ of the model, where we have a large-$N_f$ prediction
\cite{KY,kogut93}. In the scalar channel, 
the momentum space propagator
\begin{equation}
D_\sigma(k^2)\propto{{\mu^{d-2}}\over{(\surd k^2)^{d-2}+\mu^{d-2}}}
\end{equation}
where $2<d<4$ and $\mu$ is a dimensionful scale which increases as
$(g_c^2-g^2)^{1\over{d-2}}$, ie. as an
inverse correlation length. For $d=3$ this implies \cite{AbSten}
\begin{equation}
C_\sigma(t)\propto\mu\int_0^\infty dk {{\cos kt}\over{k+\mu}}\equiv\mu\int_0^\infty
d\omega {\omega\over{\omega^2+\mu^2}}e^{-\omega t},
\end{equation}
and hence the large-$N_f$ prediction
\begin{equation}
\rho_\sigma(\omega)\propto{{\mu\omega}\over{\omega^2+\mu^2}}.
\label{eq:rholargeN}
\end{equation}
In the asymptotic regime we thus have 
$\rho_\sigma\to\rho_{UV}(\omega)\propto\omega^{-1}$
rather than $\rho_0(\omega)\propto\omega$. This is a consequence of
the large non-perturbative anomalous dimension $\eta_{\bar\psi\psi}=d-2$
acquired by the scalar density at the UV fixed point \cite{kogut93},
which relates the asymptotic forms via 
\begin{equation}
\rho_{UV}(\omega)\propto\rho_0(\omega)\omega^{-2\eta_{\bar\psi\psi}}.
\end{equation}
At smaller energy scales we 
interpret $\rho$ as describing a resonance whose central
position and width are both $O(\mu)$ and hence
increase as the coupling $g^2$ is reduced. A second prediction of
(\ref{eq:rholargeN}) is that the 
dimensionless combination $\rho(\omega)/\omega$
tends to a constant in the IR limit $\omega\to0$.

Another situation of interest is the possibility of $\sigma$ decay in the
chirally broken phase. Denote the physical fermion mass by $M_f$; the
$\sigma$  is then 
expected to be a weakly bound state of mass $M_\sigma\lapprox2M_f$ whereas,
for the case of a continuous chiral symmetry, the pion mass $m_\pi$ may be much
smaller. If $2m_\pi<M_\sigma$, the decay $\sigma\to2\pi$ is allowed and
should show up as a threshold in the scalar spectral function. This should be a
good warm-up exercise for studying the physical decay $\rho\to2\pi$ in QCD; as
well as the computational saving, an important technical consideration in the
present case is that unlike in QCD
the two pions can be produced in a state of zero relative momentum.

\begin{figure}[htb]
\vskip -3cm
\begin{center}
\epsfig{file=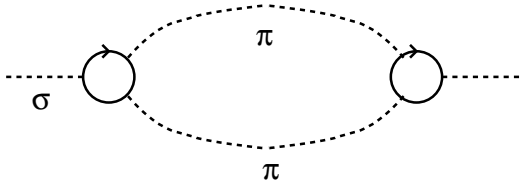, angle=-90, width=15cm}
\end{center}
\vskip -4cm
\caption[]{Contribution to $D_\sigma$ from a 2$\pi$ intermediate state.} 
\label{fig:sig2pi}
\end{figure}

Let us first derive an expectation for the form of the threshold using 
the $1/N_f$ expansion. The contribution of the two pion intermediate state to
the $\sigma$ correlator is shown diagramatically in Fig.~\ref{fig:sig2pi}.
To leading order in $1/N_f$, using the conventions of Sec.~2 of \cite{kogut93}
the $\sigma$ propagator is given by Eqn.~(\ref{eq:Dsigma})
where for momenta $k\ll M_f$ the hypergeometric function in the denominator
may be approximated by $F\approx1$.  
We will assume that for bare fermion mass $m>0$, the pion
propagator $D_\pi$ is given by the same expression with the 
factor $(k^2+4M_f^2)$ in the denominator replaced by $(k^2+m_\pi^2)$.
The vertex $\Gamma_{\sigma\pi\pi}$ is assumed to arise from a single fermion
loop as indicated in Fig.~\ref{fig:sig2pi}. It is identically zero if chiral 
symmetry is unbroken. Using the bare vertex
$-g/\surd N_f$, it is straightforward to show 
\begin{equation}
\Gamma_{\sigma\pi\pi}\simeq-G_{\sigma\pi\pi}{{g^3M_f^{d-3}}\over{\surd N_f}}
\end{equation}
where $G_{\sigma\pi\pi}$ is a dimensionless $d$-dependent constant. 

With these components in place 
it is now possible to calculate $D_\sigma$ including the 
effects of the two pion intermediate state. Specialising to $d=3$, we find
\begin{equation}
D_\sigma^{-1}(k^2\ll M_f^2)={g^2\over{4\pi M_f}}
\left[k^2+4M_f^2-{{G_{\sigma\pi\pi}^2}\over N_f}{M_f^3\over{\surd k^2}}
\tan^{-1}\left({{\surd k^2}\over{2m_\pi}}\right)\right].
\end{equation}
Besides the pole at $k^2\simeq-4M_f^2$, there is now a contribution 
at $O(1/N_f)$ to the timeslice correlation function given by
\begin{equation}
C_\sigma^{(1)}(t)
\propto{{G_{\sigma\pi\pi}^2M_f^3}\over N_f}\int{{dk}\over{2\pi}}
{{e^{ikt}}\over{k(k^2+4M_f^2)^2}}\tan^{-1}\left({k\over{2m_\pi}}\right),
\end{equation}
The two pion threshold manifests itself via branch cuts in the inverse
tangent running from $k^2=-4m_\pi^2$ out to $\pm i\infty$. Approximating
$k^2\ll M_f^2$ as before we integrate around the cut in the upper half plane 
to obtain
\begin{equation}
C_\sigma^{(1)}(t)\propto{{G_{\sigma\pi\pi}^2}\over{32N_fM_f}}
\int_{2m_\pi}^\infty{{d\omega}\over\omega}e^{-\omega t}
\end{equation}
from which we read off
\begin{equation}
\rho_\sigma^{(1)}(\omega)\propto{{G_{\sigma\pi\pi}^2}\over{32N_fM_f}}
{1\over\omega}\theta(\omega-2m_\pi).
\label{eq:rhosig2pi}
\end{equation}

Eqn.~(\ref{eq:rhosig2pi}) predicts that as well as a pole at $\omega\simeq2M_f$,
there should also be a spectral feature at $\omega=2m_\pi$ whose strength scales
as $(N_fM_fm_\pi)^{-1}$; this is in principle testable by varying the simulation
parameters $N_f$, $g^2$ and $m$.
On a finite volume it will, however, prove difficult to study the detailed form
of the spectral function above threshold. This is because the number of modes
into which the $\sigma$ can decay is strictly delimited by the allowed pion
wavevectors $\vec k_\pi=2\pi\vec n/L_s$, where $\vec n$ has integer-valued
components, and $2\sqrt{m_\pi^2+k_\pi^2}<M_\sigma$. The optical theorem,
however, implies that the only intermediate states which can contribute
to $\rho(\omega)$ are possible decay modes of the $\sigma$; 
we infer that on a finite lattice,
the $\omega^{-1}$ shape predicted by (\ref{eq:rhosig2pi}) 
is replaced by a set of 
$\delta$-functions, each arising from an allowed $\vec k_\pi$.
With imperfect (ie. finite) statistical data, however, it is possible that under
MEM these
isolated poles will blend into a continuum of approximately the correct shape.

\section{Simulations}
\label{sec:sim}

The fermionic part of the lattice action we have used for the semi-bosonized GNM$_3$
with $U(1)$ chiral symmetry is given by \cite{hands95}
\begin{eqnarray}
S_{fer} & = &   \bar{\chi}_i(x) M_{ijxy} \chi_j(y)   \nonumber \\ 
        & = &   \sum_{i=1}^{N} \left ( \sum_{x,y} \bar{\chi}_i(x) \mathcal{M}_{xy} \chi_i(y)
 +  \frac{1}{8} \sum_{x} \bar{\chi}_i(x) \chi_i(x)  
[ \sum_{ \langle \tilde{x},x \rangle}
\sigma(\tilde{x}) + i \epsilon(x) \sum_{ \langle \tilde{x},x \rangle}
\pi(\tilde{x}) ] \right ),
\end{eqnarray}
where $\chi_i$ and $\bar{\chi}_i$ are Grassmann-valued 
staggered fermion fields
defined on the lattice sites, 
the auxiliary fields $\sigma$ and $\pi$ are defined on the dual lattice
sites, 
and the symbol $\langle \tilde{x},x \rangle$ denotes the set of 8 dual lattice
sites $\tilde{x}$ surrounding the direct lattice site $x$.
The fermion kinetic operator $ \mathcal{M} $ is given by
\begin{equation}
\mathcal{M}_{x,y} = 
\frac{1}{2} \sum_{\nu} \eta_{\nu}(x) \left[ \delta_{y,x+\hat{\nu}} -
\delta_{y,x-\hat{\nu}} \right]+m\delta_{x,y},
\end{equation}
where $\eta_{\nu}(x)$ are the Kawamoto-Smit phases 
$(-1)^{x_0+\cdots+x_{\nu-1}}$,
and the symbol $\epsilon(x)$ denotes the alternating phase $(-1)^{x_0+x_1+x_2}$.
The auxiliary fields $\sigma$ and $\pi$ are weighted in the path integral by an 
additional factor corresponding to 
\begin{equation}
S_{aux}=\frac{N}{2g^{2}} \sum_{\tilde{x}} [\sigma^2(\tilde{x}) + \pi^2(\tilde{x})].
\end{equation}
The simulations were performed using a standard hybrid Monte Carlo (HMC)
algorithm
without even/odd partioning, implying that 
simulation of $N$ staggered
fermions describes $N_f=4N$ continuum species \cite{hands95}; 
the full symmetry of the lattice model in the
continuum limit, however, is $U(N_f/2)_V \otimes U(N_f/2)_V \otimes U(1)$ 
rather than $U(N_f)_V \otimes U(1)$. 
At non-zero lattice spacing the symmetry group is smaller still: 
$U(N_f/4)_V \otimes
U(N_f/4)_V \otimes U(1)$. 
In the  $Z_2$-symmetric model the $\pi$ fields are switched off and $M$ 
becomes real so that real pseudofermion fields can be used. In this case 
$N$ staggered fermions describe $N_f=2N$ continuum species.
Further details of the algorithm and the optimisation of 
its performance can be found in \cite{kogut93,hands95}.

Using point sources we calculated the zero momentum fermion $(f)$ correlator
at different values of the coupling $\beta\equiv1/g^2$. In order to compare 
MEM to conventional spectroscopy we also estimated the fermion mass using a
simple pole fit using the function
\begin{equation}
C_f(t)=A_f [e^{-M_ft}-(-1)^t e^{-M_f(L_t-t)}].
\end{equation}
Similarly, the zero momentum auxiliary $\pi$ correlator was measured and its
mass estimated using a cosh fit.
The mesonic correlators are given by:
\begin{equation}
C_{M}(t)=\sum_{\mathbf{x},\bf{x}_1,\mathbf{x}_2} \Phi(\mathbf{x}_1) \Phi(\mathbf{x}_2)
W_{M}({\mathbf x}) G(\mathbf{x},t;\mathbf{x}_1,0)
G^{\dagger}(\mathbf{x},t;\mathbf{x}_2,0),
\label{eq:meson}
\end{equation}
where $G$ is the lattice fermion propagator and $W_{M}({\mathbf x})$ 
a 
phase factor which picks out a channel with particular
symmetry properties i.e. $W_{M}({\mathbf x}) = \epsilon(x)$ 
for the S channel and 
$W_{M}({\mathbf x}) = 1$ for the PS channel.
The function $\Phi(\mathbf{x})$ is either a point source
$\delta_{\mathbf{x},(0,0)}$ 
or a staggered fermion wall source $\sum_{m,n=0}^{L_s/2-1}
\delta_{\mathbf{x},(2m,2n)}$ \cite{gupta}.
In all the simulations we used point sinks.
These correlators were fitted to a function $C_{M}(t)$ given by
\begin{equation}
C_M(t)=A[ e^{-M_Mt} + e^{-M_M(L_t-t)}] + \tilde{A} (-1)^t [e^{-\tilde{M}_Mt} 
+ e^{-\tilde{M}_M(L_t-t)}].
\end{equation}
Note that composite operators made from staggered fermion
fields project onto more than one set of continuum quantum numbers.
The first square bracket represents the ``direct'' signal with mass $M_M$ 
and the second an ``alternating'' signal
with mass $\tilde{M}_M$. Continuum quantum numbers for various mesonic channels
are given in \cite{HKS} -- in this study we focus on the PS$_{direct}$ channel, 
with $J^P=0^-$. Although expected to be the tightest bound meson 
since it is the
only one for which $s$-wave binding is available, as stressed in 
\cite{hands95,HKS} this state does {\sl not\/} project onto the Goldstone mode
in the broken phase.

\section{Results}
\label{sec:results}
\subsection{The $\pi$, $f$ and PS Channels in the Broken Phase}

\begin{figure}[b]
\begin{center}
\epsfig{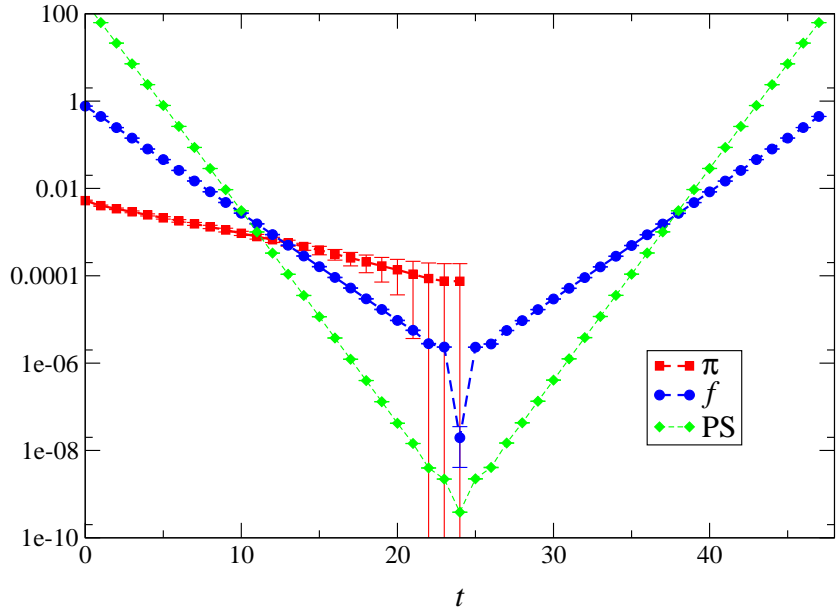}
\end{center}
\caption{
Propagators in three different channels from
simulations of the U(1) model on a $32^2\times48$ lattice at $\beta=0.55$,
$m=0.01$.}
\label{fig:corr_broken}
\end{figure}
\begin{figure}[hbt]
\begin{center}
\epsfig{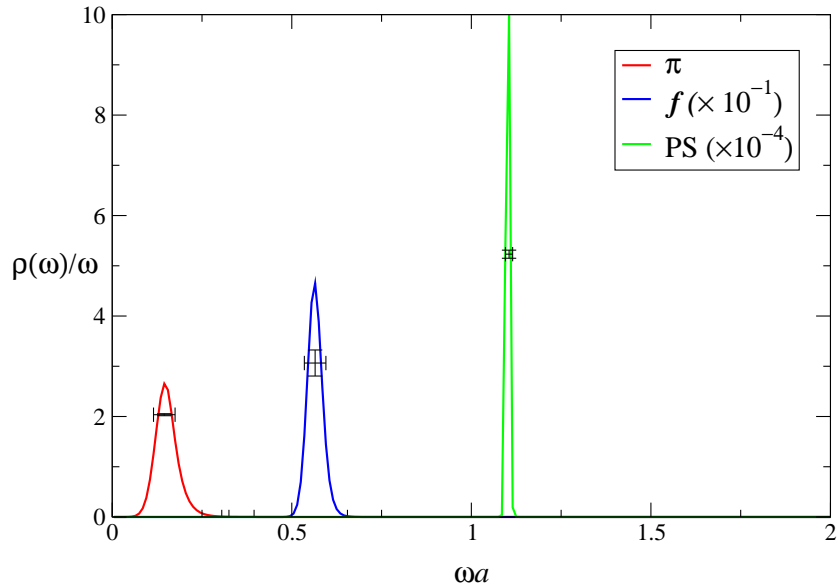}
\end{center}
\caption{
Bryan image of $\rho(\omega)/\omega$ in three different channels 
using the same data as Fig.~\ref{fig:corr_broken}.}
\label{fig:allparts}
\end{figure}
We first discuss results from the chirally broken phase, obtained 
with $\beta<\beta_c\approx1.0$. Fig.~\ref{fig:corr_broken} shows the
propagators for $\pi$, $f$ and PS channels on a log scale 
(using data obtained with a wall source and point sink
in the latter
case), resulting from approximately 40000 HMC trajectories of mean length 1.0. 
All three look to be well-approximated by straight lines, implying that
each channel is dominated by a single particle pole.
Fig.~\ref{fig:allparts} shows the spectral functions obtained in
the same three channels 
using Bryan's method. All three appear as
well-localised peaks suggesting simple poles and hence stable particle
states. The cross shown on each peak is obtained as follows.
The spectral feature is fitted to a form $ZG(\omega-M/\Gamma)$ where
$G(x)$ is the normalised gaussian distribution, $M$ the peak position,
$\Gamma$ the full width at half maximum, and $Z$ a normalisation factor.
The horizontal bar's position and width represent $M$ and $\Gamma$ respectively,
and its height represents the area of $ZG(\omega-M/\Gamma)$ evaluated
between $\omega-\Gamma$ and $\omega+\Gamma$. The vertical error bar represents 
the error in this area as determined by the Bryan algorithm \cite{MEM.90.RKB}.
For a narrow 
gaussian, of course, the central value is interpreted as the particle mass. 

\begin{table}[b]
\centering
\begin{tabular}{|ccccllll|}
\hline
& $N_f$ & Volume & $\beta$ & $m$ & Mass & Mass & Area \cr
&       &        &         &     &(1-exp)&(MEM)&      \cr
\hline
\hline
$\pi$&4     & $32^2\times48$ & 0.55    & 0.005 & 0.114(4)  & 0.112(6)  &
0.501(129)
 \cr
&4     & $32^2\times48$ & 0.55    & 0.01  & 0.168(5)  & 0.154(9)  &
0.176(15)
 \cr
&4     & $32^2\times48$ & 0.55    & 0.02  & 0.232(5)  & 0.231(7)  &
0.0617(98)
 \cr
&4     & $32^2\times48$ & 0.55    & 0.03  & 0.280(10) & 0.263(15) &
0.0351(37)
 \cr
&4     & $32^2\times48$ & 0.55    & 0.045 & 0.349(8)  & 0.326(14) &
0.0193(15)
 \cr
&4     & $32^2\times48$ & 0.55    & 0.06  & 0.447(24) & 0.435(1.9)&
0.0102(5.7)
\cr
&4     & $32^2\times48$ & 0.65    & 0.01  & 0.193(4)  & 0.187(8)  &
0.0810(78)
 \cr
&4     & $32^2\times48$ & 0.65    & 0.02  & 0.277(4)  & 0.267(6)  &
0.0289(19)
 \cr
&36    & $24^2\times32$ & 0.55    & 0.01  & 0.150(5)  & 0.144(18) &
0.053(19)
 \cr
&36    & $24^2\times32$ & 0.55    & 0.02  & 0.238(6)  & 0.229(8)  &
0.0140(14)
 \cr
&36    & $24^2\times32$ & 0.55    & 0.03  & 0.287(10) & 0.271(17) &
0.0081(10)
 \cr
\hline
\hline
$f$&4     & $32^2\times48$ & 0.55    & 0.005 & 0.555(7)  & 0.556(4)   &
2.15(49)
\\
&4     & $32^2\times48$ & 0.55    & 0.01  & 0.564(1)  & 0.564(1)   & 2.37(3)
\\
&4     & $32^2\times48$ & 0.55    & 0.02  & 0.5853(7) & 0.5858(13) &
2.14(27)
\\
&4     & $32^2\times48$ & 0.55    & 0.03  & 0.599(1)  & 0.599(1)   & 2.06(5)
\\
&4     & $32^2\times48$ & 0.55    & 0.045 & 0.623(1)  & 0.623(1)   & 1.90(4)
\\
&4     & $32^2\times48$ & 0.55    & 0.06  & 0.644(2)  & 0.643(2)   & 1.63(8)
\\
&4     & $32^2\times48$ & 0.65    & 0.01  & 0.3978(8) & 0.3965(13) & 5.11(9)
\\
&4     & $32^2\times48$ & 0.65    & 0.02  & 0.4285(6) & 0.4384(44) &
4.10(33)
\\
&36    & $24^2\times32$ & 0.55    & 0.01  & 0.6796(3) & 0.6796(3)  & 1.77(8)
\\
&36    & $24^2\times32$ & 0.55    & 0.02  & 0.6911(3) & 0.6908(3)  & 1.72(7)
\\
&36    & $24^2\times32$ & 0.55    & 0.03  & 0.7025(4) & 0.7023(5)  & 1.59(2)
\\
\hline
\hline
PS& 4     & $32^2\times48$ & 0.55    & 0.005 & 1.0807(8)  & 1.0807(8)   &
164.3(6)
    \\
& 4     & $32^2\times48$ & 0.55    & 0.01  & 1.0973(8)  & 1.0979(7)   &
160(3)
     \\
& 4     & $32^2\times48$ & 0.55    & 0.02  & 1.1395(6)  & 1.1396(5)   &
147.2(5)
    \\
& 4     & $32^2\times48$ & 0.55    & 0.03  & 1.1715(11) & 1.1716(11)  &
130(2)
     \\
& 4     & $32^2\times48$ & 0.55    & 0.045 & 1.2253(6)  & 1.2231(6)   &
119.1(9)
    \\
& 4     & $32^2\times48$ & 0.55    & 0.06  & 1.2693(13) & 1.2691(2)   &
103(2)
     \\
& 4     & $32^2\times48$ & 0.65    & 0.01  & 0.7722(6)  & 0.7711(4)   &
426(32)
     \\
& 4     & $32^2\times48$ & 0.65    & 0.02  & 0.8362(5)  & 0.8381(45)  &
343(462)
    \\
& 36    & $24^2\times32$ & 0.55    & 0.01  & 1.3568(2)  & 1.3569(2)   &
50.1(3)
     \\
& 36    & $24^2\times32$ & 0.55    & 0.02  & 1.3806(2)  & 1.3808(2)   &
48.4(2)
     \\
& 36    & $24^2\times32$ & 0.55    & 0.03  & 1.4030(3)  & 1.4030(3)   &
45.5(3)
     \\
\hline
\end{tabular}
\caption{Broken phase spectroscopy}
\label{tab:spect}
\end{table}
In Table~\ref{tab:spect} we list the masses obtained from simulations
of the U(1) model from both 
single exponential fits and MEM, as well as the area under the gaussian peak, 
using correlator data from timeslices 2 -- 10 
for the $\pi$; for $f$ and PS timeslices 2 -- 8 were used. Note that 
for the lightest state, namely the $\pi$, MEM systematically yields a 
lower mass, suggesting that it is less affected by 
excited state contamination,
although in all cases the two methods are within a standard deviation.
\begin{figure}[htb]
\begin{center}
\epsfig{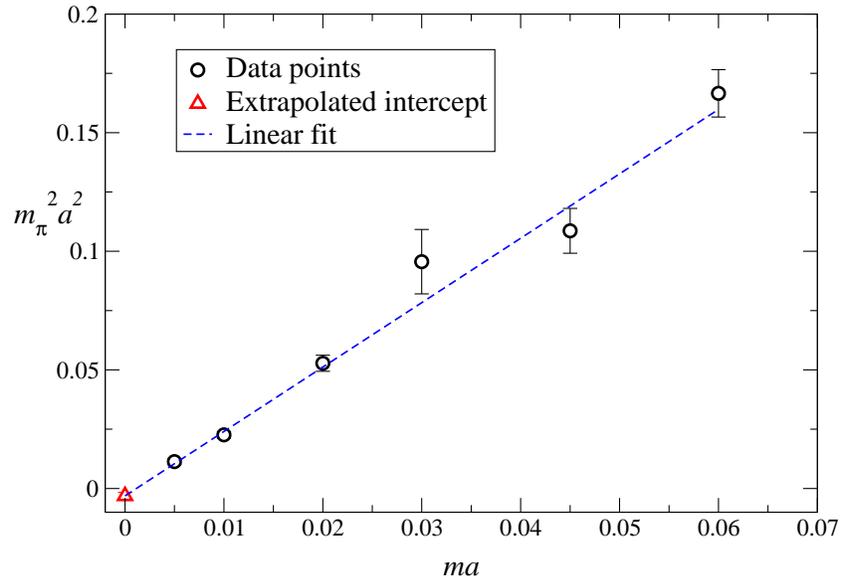}
\end{center}
\caption{Pion mass $m_\pi^2$ vs. bare mass $m$ for $\beta=0.55$, showing
evidence for the Goldstone nature of the $\pi$.}
\label{fig:pcac}
\end{figure}
\begin{figure}[htb]
\begin{center}
\epsfig{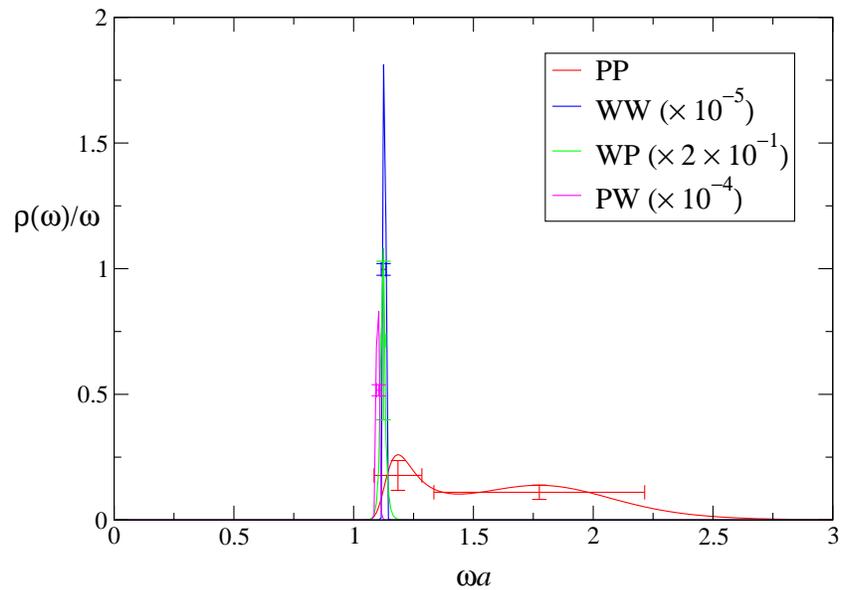}
\end{center}
\caption{
Bryan image of $\rho(\omega)/\omega$ in the PS channel
$32^2\times48$ lattice at $\beta=0.55$, $m=0.01$ using correlators 
with different combinations of wall and point sources.}
\label{fig:allsources}
\end{figure}
Fig.~\ref{fig:pcac} demonstrates
that the pion mass extracted using MEM over a range 
of bare fermion masses is consistent with the PCAC 
behaviour $m_\pi\propto\surd m$
expected for broken chiral symmetry.
For the $f$ and PS channels 
there is excellent agreement in almost all cases between the two methods.
The PS mass is roughly twice that of the fermion, consistent with its being a 
weakly bound state.
With the precision we have obtained it is possible to estimate the binding
energy defined as $\Delta_M=2M_f-M_{\rm PS}$; the results are tabulated in
Table~\ref{tab:be}.
\begin{table}[!htb]
\vskip 5 truecm
\centering
\begin{tabular}{|ccclll|}
\hline
$N_f$ & Volume   & $\beta$ & $m$ & $\Delta_M$ & $\Delta_M$ \\
      &          &         &     & (1-exp)    & (MEM)   \\
\hline
4     & $32^2\times48$ & 0.55    & 0.005 & 0.0293 & 0.0313   \\
4     & $32^2\times48$ & 0.55    & 0.01  & 0.0307 & 0.0301   \\
4     & $32^2\times48$ & 0.55    & 0.02  & 0.0311 & 0.0320   \\
4     & $32^2\times48$ & 0.55    & 0.03  & 0.0265 & 0.0264   \\
4     & $32^2\times48$ & 0.55    & 0.045 & 0.0207 & 0.0229   \\
4     & $32^2\times48$ & 0.55    & 0.06  & 0.0187 & 0.0169   \\
4     & $32^2\times48$ & 0.65    & 0.01  & 0.0234 & 0.0219   \\
4     & $32^2\times48$ & 0.65    & 0.02  & 0.0208 & 0.0387   \\
36    & $24^2\times32$ & 0.55    & 0.01  & 0.0024 & 0.0023   \\
36    & $24^2\times32$ & 0.55    & 0.02  & 0.0016 & 0.0008   \\
36    & $24^2\times32$ & 0.55    & 0.03  & 0.0020 & 0.0016   \\
\hline
\end{tabular}
\caption{Binding Energy in the PS channel}
\label{tab:be}
\vskip 5 truecm
\end{table}
For $N_f=4$ $\Delta_M\approx2.8\%$ of the bound state mass, but the figure drops
to $\approx0.15\%$ for $N_f=36$, which is roughly consistent with the 
analytical expectation that $\Delta_M\propto1/N_f$ (note, however, that the 
$N_f=36$ simulations were performed on a smaller volume). 
It was observed in
\cite{HKS} that the PS wavefunction has considerably greater spatial extent 
for larger $N_f$, again implying it is less strongly bound.

As discussed in Sec.~\ref{sec:theory}
the area under the peak is related to the strength $A$ of the coupling of the 
operator $J$ to the single particle state, and hence 
to physical decay constants. Our results show a systematic
decrease in this coupling
strength with bare fermion mass $m$, the effect being most pronounced for 
the $\pi$. 

Finally in Fig.~\ref{fig:allsources} we explore the effects of using different
meson sources following Eqn. (\ref{eq:meson}) 
using data from timeslices 1 -- 8. 
As in Fig.~\ref{fig:allparts}, the
spectral functions have been rescaled so as all to fit on the same plot.
When a wall is used at either sink or source, 
the signal is completely dominated
by the bound state; however, for the point-to-point correlator there is a
significant contribution out to $\omega a\approx2.5$. Since we have discarded
data from small timeslices we should not expect much quantitative 
information from the asymptotic form of $\rho(\omega)$ in this 
case; indeed, as
$\omega\to\infty$ it decays much faster than either of the 
idealised forms $\rho_0(\omega)$ or $\rho_{UV}(\omega)$
discussed in Sec.~\ref{sec:theory}.
Fig.~\ref{fig:allsources} provides a graphic illustration, however, 
of the importance of
choice of source in maximising the projection onto the ground state.
 
\subsection{Symmetric Phase}

\begin{figure}[htb]
\begin{center}
\epsfig{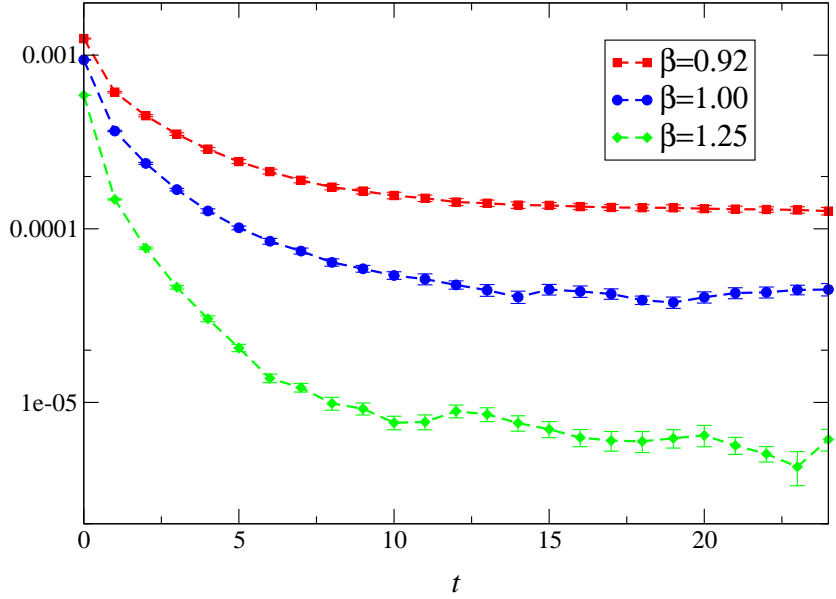}
\end{center}
\caption{$\sigma$ correlator for 3 different couplings
in the chirally symmetric phase 
on a $32^2\times48$
lattice.}
\label{fig:corr_sym}
\end{figure}
\begin{figure}[htb]
\begin{center}
\epsfig{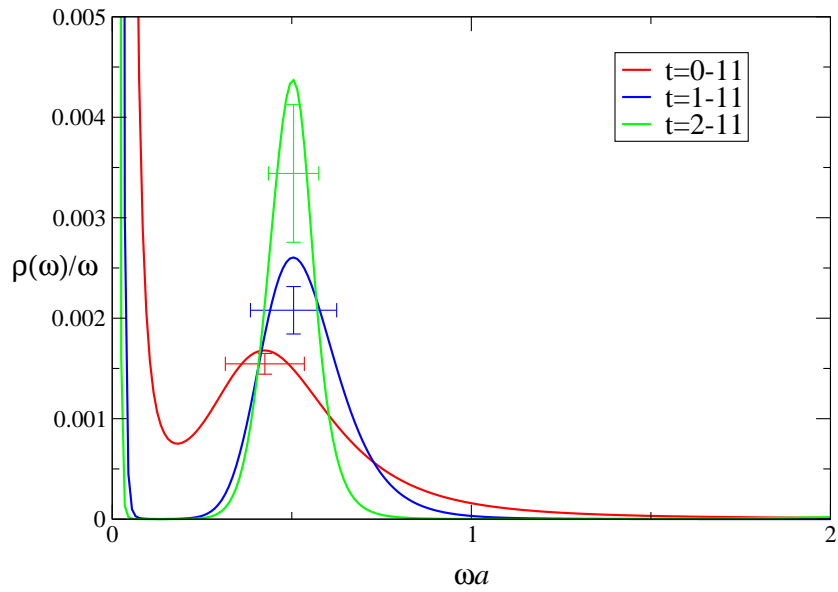}
\end{center}
\caption{Bryan image of $\rho(\omega)/\omega$ vs. 
$\omega$ in the $\sigma$ channel at $\beta=1.25$ on a $32^2\times48$
lattice, showing 3 different time windows.}
\label{fig:sym_varwindow}
\end{figure}
\begin{figure}[htb]
\begin{center}
\epsfig{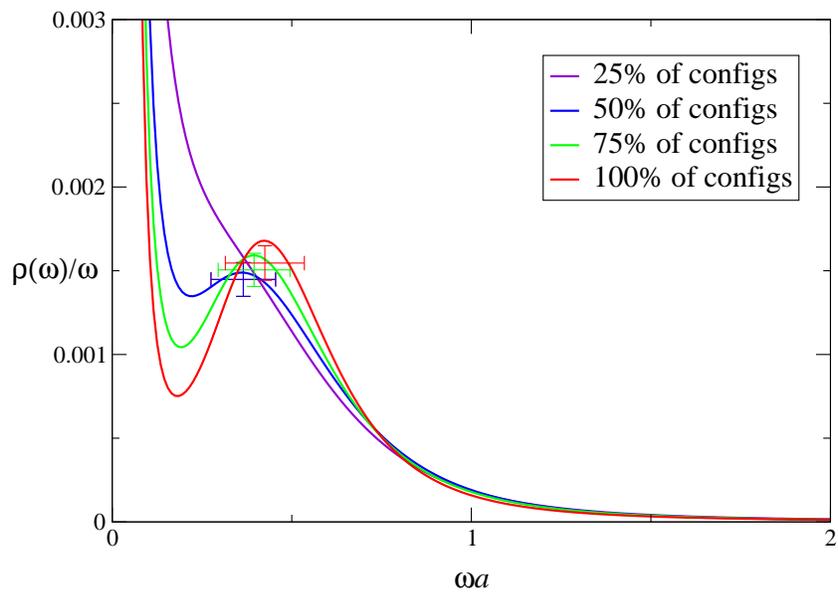}
\end{center}
\caption{The same as Fig.~\ref{fig:sym_varwindow}
using fits from timeslices 0 -- 11,
showing the effects of varying the amount of data.}
\label{fig:sym_varncfg}
\end{figure}
Next we turn to the chirally symmetric phase found for $\beta>\beta_c$, where
according to the discussion of Sec.~\ref{sec:theory} the bound state poles
should be replaced by resonances with non-vanishing widths. Our simulations
in this section were performed for the Z$_2$ model
on a $32^2\times48$ 
lattice at
couplings $\beta=0.92$, 1.0 and 1.25 with $O(40000)$ configurations
separated by HMC trajectories of mean length 1.0, 
and for U(1) on a $32^3$ lattice 
at $\beta=1.0$ and 1.25 with respectively 30000 and 60000 trajectories of mean
length 0.6. In all cases $N_f=4$ fermion flavors were used. 
It proved considerably easier in this
phase to simulate 
the model with Z$_2$ chiral symmetry: 
the U(1) simulations
required a much smaller molecular dynamics timestep 
making them more expensive, and the data correspondingly of
not such good quality. Data for the Z$_2$ $\sigma$ 
timeslice correlator are shown 
on a log scale in Fig.~\ref{fig:corr_sym}. In contrast to the broken phase
correlators of Fig.~\ref{fig:corr_broken} it is clear that a simple pole fit
will not be successful; indeed, the correlators become almost flat at large
$t$, which means that towards the centre of the lattice we have to worry 
about significant contributions from not just
a backwards-propagating signal, but also 
``image''
sources displaced by integer mulitples of $L_t$ from the original source
\cite{kogut93}.

If we are to
successfully identify spectral features as something other than
simple poles, then it is
important to study systematic effects. Fig.~\ref{fig:sym_varwindow} 
presents
results from the $\sigma$ channel, where the resonance is anticipated,
showing
the effects of varying the timeslice sample used in the MEM fit. 
Data from within a time window $[t_1,t_2]$ were fitted; in all cases we chose
a rather conservative value $t_2=11$
to minimise finite volume (actually non-zero
temperature) effects due to the image sources discussed above, although
we have checked that the results are insensitive to reducing $t_2$.
Fig.~\ref{fig:sym_varwindow} shows a broad feature centred at 
$\omega a\simeq0.5$, whose ``width'' (actually the ratio of width to area, as
indicated by the crosses) increases as data from smaller times is included.
Ignoring the divergence as $\omega\to0$ which we take to be an artifact 
(possibly due to a small residual vacuum expectation
$\langle\sigma\rangle\not=0$; see discussion below in Sec.~\ref{sec:sigma}),
the shape of the spectrum appears qualitatively similar to the
large-$N_f$ prediction (\ref{eq:rholargeN}).
The fact that the shape of the spectrum in the massless phase is sensitive to
the data at short times is slightly counter-intuitive, but is consistent with 
the observation in \cite{kogut93} that extraction of a physical scale, namely
the resonance width $\mu$,
from timeslice correlator data actually depends on corrections to the 
expected power-law
falloff $(\mu t)^{-2}$ at {\em small} values of $\mu t$.
Note that $\rho(\omega)/\omega$ falls to zero as $\omega\to\infty$, 
in contrast to the constant behaviour expected in an asymptotically-free theory 
such as QCD and exemplified in Fig.~\ref{fig:testing_mem}. The falloff is 
approximately power-law of the form $\omega^{-p}$, but with $p\approx4 -
6$, in contrast to the value $p=2$ predicted by (\ref{eq:rholargeN}).

It is also legitimate to ask whether the non-zero width of the spectral feature
is due to insufficient statistics. Fig.~\ref{fig:sym_varncfg} shows the feature 
evolving as data is added to the sample. There is no significant reduction in
the width of the feature as the statistics accumulate from $O(10000)$ to
$O(40000)$ configurations, although the central position and height of the peak
both vary slightly, supporting the conclusion that a resonance is present.

\begin{figure}[htb]
\begin{center}
\epsfig{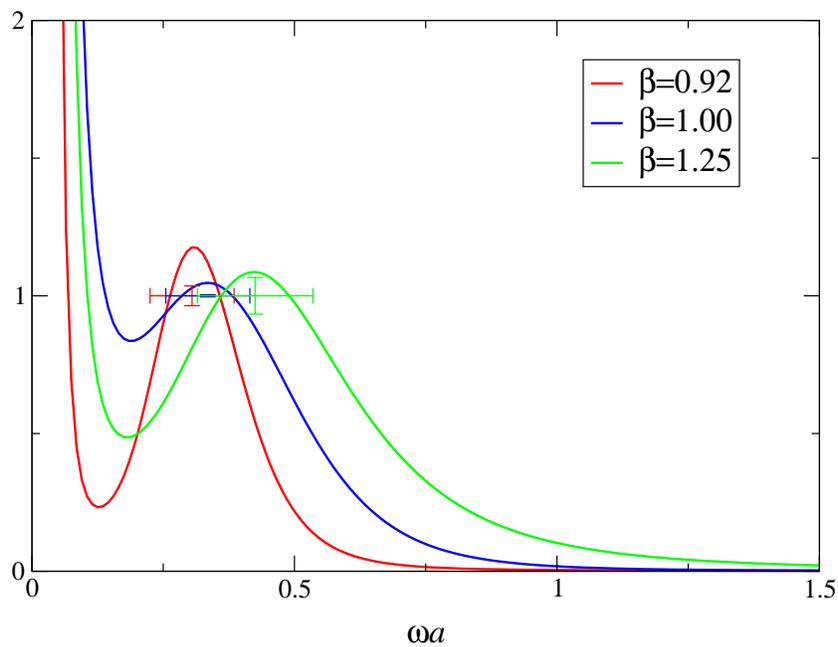}
\end{center}
\caption{Rescaled Bryan image of $\rho(\omega)/\omega$ in the $\sigma$ channel
from timeslices 0 -- 11,
for three different couplings.}
\label{fig:sym_3betas}
\end{figure}
\begin{figure}[b]
\begin{center}
\vskip 0.5cm
\epsfig{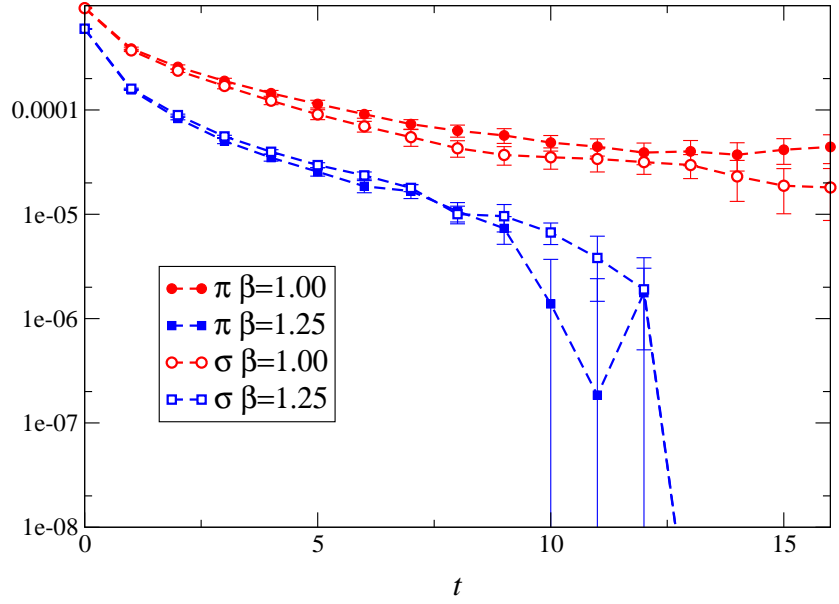}
\end{center}
\caption{$\sigma$ and $\pi$ timeslice correlators
from simulations of the U(1) model on a $32^3$ lattice.}
\label{fig:corr_symZU1}
\end{figure}
\begin{figure}[b]
\begin{center}
\vskip 0.5cm
\epsfig{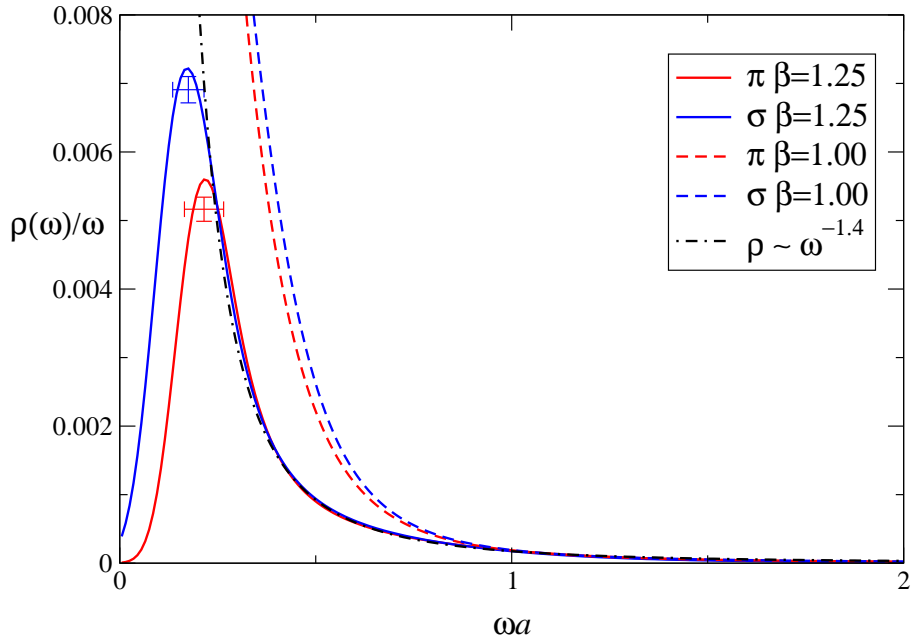}
\end{center}
\caption{Bryan image of $\rho(\omega)/\omega$ in both $\sigma$ and $\pi$
channels using the correlator data of Fig.~\ref{fig:corr_symZU1}.}
\label{fig:sym_pi_sig}
\end{figure}
In Fig.~\ref{fig:sym_3betas} we compare the results from 3 different couplings.
Since the artifact at $\omega\to0$ distorts the normalisation of our result, we
have rescaled each curve so that the rectangles of equal area to the fitted peak
have the same height. The resulting curves show both the position and width of
the resonance increasing with $\beta$. This is consistent with
(\ref{eq:rholargeN}), which predicts both are proportional to a 
single scale $\mu$, if $\mu$ 
increases with $\beta$ as expected. Within errors we find 
the ratio of width to central
position constant and approximately equal to 50\%. Note, however, that ignoring
the spike at $\omega=0$ the dip in the curve suggests
$\lim_{\omega\to0}\rho(\omega)/\omega\to0$, rather than tending to a constant
as predicted by (\ref{eq:rholargeN}).

Finally in Figs.~\ref{fig:corr_symZU1} and \ref{fig:sym_pi_sig}
we show some results from simulations of
the U(1) model. 
In this case it is possible to extract and compare
spectra from both $\sigma$ and $\pi$ channels.
The fitted time window is [1,10].
The bare fermion mass $m$ is set to zero implying that for
$\beta>\beta_c$ the two channels should be physically indistinguishable, and 
Fig.~\ref{fig:sym_pi_sig} suggests that for large $\omega$ this is indeed the
case. There is, however, a large disparity as $\omega\to0$ between
$\beta=1.00$, where $\rho(\omega)$ appears to diverge, and $\beta=1.25$ where
it seems to tend smoothly to zero. Fig.~\ref{fig:corr_symZU1} confirms that
the behaviour of the correlators at large $t$ is not really under control yet
with
the precision we have been able to obtain.
Also in both cases there is more power in the
$\sigma$ channel at small $\omega$. This indicates we still lack a full
understanding of systematics in this regime. Intriguingly, however, the
large-$\omega$ behaviour is much closer to the large-$N_f$ prediction;
the dash-dotted line in Fig.~\ref{fig:sym_pi_sig} 
is a fit of the form $\rho(\omega)\propto\omega^{-1.4}$, to be compared
with the expected $\omega^{-1}$.

To summarise, there is encouraging evidence 
that MEM analysis can successfully
identify a resonance with non-zero width in this phase of the model, whose
properties are consistent, at least in part, with theoretical 
expectations. 
Despite uncertainty about the $\omega\to0$ limit that would probably
require lattices
considerably longer in the Euclidean time dimension to resolve, the MEM 
method is capable of yielding semi-quantitative information in this regime.

\subsection{The $\sigma$ Channel in the Broken Phase}
\label{sec:sigma}

\begin{figure}[htb]
\begin{center}
\epsfig{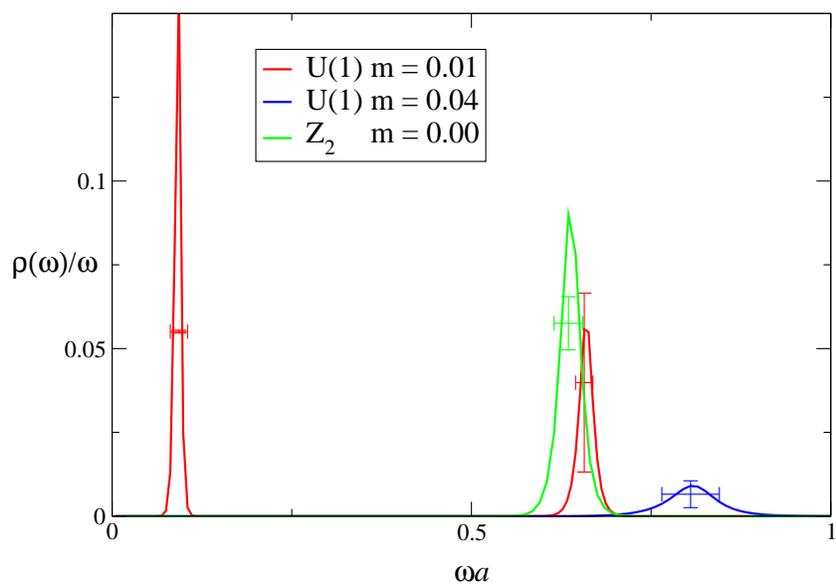}
\end{center}
\caption{Rescaled Bryan image of $\rho(\omega)/\omega$ in the $\sigma$ channel
from timeslices 1 -- 10
at $\beta=0.65$, for two different masses in the U(1) model on a $32^2\times24$
lattice, and for $m=0$ in the Z$_2$ model on a $24^3$ lattice.} 
\label{fig:sigma}
\end{figure}
Finally we return to the chirally broken phase and switch our attention to the
$\sigma$ channel. Recall that since the $\sigma$ is modelled via an auxiliary 
boson field, diagrams formed from disconnected
fermion lines are automatically included in the calculation of the correlator.
The main physical issues to address are whether the the $\sigma$ is a
bound state, and if it is possible to detect a signal for $\sigma\to\pi\pi$
decay. Conventional spectroscopy, using both simple pole fits and large-$N_f$
inspired forms which include a $f\bar f$-threshold, have proved at best
ambiguous for this case \cite{kogut93}. Moreover due to the auxiliary nature of
the field it is impossible to study the wavefunction, which in other channels 
provides clear evidence of $f\bar f$ binding \cite{HKS}.

Fig.~\ref{fig:sigma} shows spectral functions in the $\sigma$ channel from 
simulations of the U(1) model at two different values of bare fermion mass $m$,
and a comparison simulation of the massless Z$_2$ model, in which there is no
pion degree of freedom. We used a large statistical sample; respectively
$1.7\times10^6$ (U(1) $ma=0.01$),
$4\times10^5$ (U(1) $ma=0.04$), and $1.1\times10^6$ (Z$_2$ $m=0$)
configurations were generated, and
in all cases $N_f=4$. 
Since the $\sigma$ has the same quantum numbers as the
vacuum, it is necessary to subtract the vacuum term
$C_{vac}=\sum_{\vec x,t}\langle\sigma(\vec0,0)\rangle\langle\sigma
(\vec x,t)\rangle$ from the raw data 
to define a connected Green function. 
Because of statistical fluctuations
this procedure is hard to implement exactly, despite the large sample generated,
and we believe that uncertainity in the vacuum subtraction 
is the origin
of the sharp spike in the U(1) $m=0.01$ spectrum centred at 
$\omega a=0.092$. This feature is otherwise hard to explain since 
the lightest particle in the spectrum (see Table~\ref{tab:spect}), the $\pi$,
has mass $m_\pi a\simeq0.19$. We have checked that varying the subtraction 
constant $C_{vac}$ within a standard deviation causes dramatic alterations to 
both the strength and position of this feature without significantly
affecting the peaks at higher $\omega$, and conclude that it is not
physical. 

Proceeding on this assumption we identify spectral features centred at
$\omega a=0.81(2)$ (U(1) $ma=0.04$), $\omega a=0.66(1)$ (U(1) $ma=0.01$), 
and
$\omega a=0.64(1)$ (Z$_2$ $ma=0$). 
The width of the features are $O(0.05)$ and appear 
stable as the number of
configurations sampled is increased, which suggests they are not simple
poles. Unlike the PS spectrum of Fig.~\ref{fig:allsources},
however, their shapes are roughly symmetric, which contrasts with the 
large-$N_f$ expectation that $\rho(\omega)$ should be sharply cut off on the
low-$\omega$ side but fall away more slowly on the high-$\omega$ side due to a 
$f\bar f$ continuum. The central value of the peak for the U(1)
$m=0.01$ data indicates that the 
state it describes is lighter than the corresponding PS state
in the U(1) model (see Table~\ref{tab:spect}), which has mass 0.77 -- 
the $f\bar f$ threshold in this case is at 0.793(3), which lies well above
the point where $\rho(\omega)/\omega$ appears to fall to zero. We deduce that
for finite $N_f$ there is a bound state in the $\sigma$ channel, which 
is more tightly-bound than the PS meson for which there are
no disconnected fermion line contributions. This conclusion
would have been difficult to reach without MEM.

Unfortunately there is no sign of any spectral feature at the two pion
threshold, expected following the discussion of Sec.~\ref{sec:theory}
at $\omega a\simeq0.38$ for $ma=0.01$ and $\omega a\simeq0.75$
for $ma=0.04$ 
(implying that $\sigma\to\pi\pi$ decay is certainly possible on 
energetic grounds in the
former case). We have checked that there is no significant difference between 
spectra found in U(1) and Z$_2$ simulations performed at the same parameter
values. Possibly this is because the height of the expected feature is 
suppressed by a power of $1/N_f$ (recall (\ref{eq:rhosig2pi})) and would need a
series of simulations with varying $N_f$ to expose it. Thus far, however, 
we are unable to report observation of bound state
decay in this model.

\section{Summary and Outlook}
\label{sec:summary}

Lattice simulations of theories other than quenched QCD at zero temperature 
will require spectrum analysis techniques of greater sophistication than the 
currently-used single- and multi-exponential fits, which implicitly assume a
spectral density function made up from a series of isolated simple poles. In
this paper we have applied one of the more promising, the Maximum Entropy
Method, for the first time to a lattice model with dynamical fermions.
Our main findings are summarised below:

\begin{itemize}

\item
In the chirally broken phase of the model we have found sharply defined spectral
features corresponding to the elementary fermion $f$, 
the simplest mesonic $f\bar f$
bound state, and the Goldstone boson $\pi$. These results corroborate earlier 
simulations \cite{hands95,HKS}, and for the first time have permitted
a plausible estimate for the meson binding energy.

\item
In the chirally symmetric phase we have identified a broad resonance feature
whose position and width agree qualitatively with the expectations of the
large-$N_f$ approach. The behaviour as $\omega\to\infty$ is distinct 
from that of 
an asymptotically-free theory, and is evidence for a non-perturbative 
anomalous dimension associated with a UV renormalisation group fixed point.

\item
In the chirally broken phase we have made the first quantitative
study of the $\sigma$
channel, and found that it is more tightly bound than the 
conventional PS meson, possibly
due to the additional contribution of disconnected fermion line diagrams.
We have been unable to find evidence for $\sigma\to\pi\pi$ decay.

\end{itemize}

Since the philosophy of the MEM method is to make the maximum possible use of
data, 
we have used correlator data from as wide a time window as
possible consistent with stability of the fit.
The main problem we have faced has been systematic errors associated with 
the upper end of the time window used in the fit, particularly since we have
been anxious to avoid finite volume effects. This has made it impossible to 
have control of the $\omega\to0$ limit. As explained in
Sec.~\ref{sec:sigma}, in the $\sigma$ channel 
there may also be artifacts associated 
with vacuum subtraction. 
Overall, our conclusion is that MEM has proved a useful
semi-quantitative analysis tool, but that there remains room for improvement.

In the future it will be interesting to study 2+1$d$ four-fermion
models at non-zero
temperature and/or density. Since spectral analysis requires data from many 
Euclidean time separations to be effective, it is likely to be some time before
an equivalent analysis can be applied to QCD with dynamical
fermions\footnote{In this context it would be interesting to explore four-fermi
models using anisotropic lattices.}. However, in the vicinity of the 
deconfining/chiral symmetry restoring transition 
the dominant modification to the zero-$T$ spectrum is
expected to be collision-broadening due to pions, an effect 
absent from quenched QCD, where the lightest states are glueballs, but in
principle present in the current model. Additionally, there is no longer any
ambiguity about the IR cutoff, which is now $T^{-1}$, and the $\omega\to0$
limit should become accessible \cite{FLPSW}; 
the slope of $\rho(\omega)$ in this
limit yields information about transport coefficients such as
electrical conductivity via the
Kubo formula \cite{ABR}. Finally, the four-fermi model is currently the only 
model simulable at non-zero baryon chemical potential $\mu$ which has a Fermi
surface \cite{general}; there may be rich physics associated with 
phenomena such as 
first and zero sound or superfluidity via BCS condensation to explore.  

\section*{Acknowledgements}
SJH and CGS were supported by the Leverhulme Trust.
SJH thanks ITP, Santa Barbara 
(supported by the National Science 
Foundation under Grant No. PHY99-07949) and ECT$^*$, Trento  
for hospitality
during the latter stages of this work.
JBK was supported in part by NSF grant NSF-PHY-0102409.
The computer simulations were done on the Cray SV1's at NERSC, the Cray T90
at NPACI, 
and on the SGI Origin 2000 at the University of Wales Swansea.  
We are also grateful for fruitful discussions with Frithjof Karsch, 
Manfred Oevers, and Ines Wetzorke.



\end{document}